\documentstyle[aps,floats,epsf]{revtex}  

\draft
\begin{document}

\twocolumn[\hsize\textwidth\columnwidth\hsize\csname
@twocolumnfalse\endcsname

\title{Electric-field induced polarization paths in Pb(Zr$_{1-x}$Ti$_{x}$)O$_{3}$ alloys}
 
\author{L. Bellaiche$^{1}$, Alberto Garc\'{\i}a$^{2}$ and David Vanderbilt$^{3}$}

\address{
$^{1}$Physics Department, University of Arkansas, Fayetteville,
      Arkansas 72701, USA\\
$^{2}$Departamento de Fisica Aplicada II, Universidad del Pais Vasco,
      Apartado 644, 48080 Bilbao, Spain \\
$^{3}$Center for Materials Theory, Department of Physics and Astronomy,
      Rutgers University,\\ Piscataway, New Jersey 08855-0849, USA}

\date{April 6, 2001}

\maketitle
\begin{abstract}

Properties of Pb(Zr$_{1-x}$Ti$_{x}$)O$_{3}$ (PZT) for compositions
$x$ near the morphotropic phase boundary and under an electric
field are simulated using an {\it ab-initio} based approach.
Applying an electric field of  {[111]} orientation to tetragonal PZT
(e.g., $x$=0.50) leads to the expected sequence of tetragonal,
A-type monoclinic, and rhombohedral structures.  However, the
application of a field of orientation {[001]} to rhombohedral PZT
(e.g., $x$=0.47) does not simply reverse this sequence.  Instead,
the system follows a complicated path involving also triclinic
and C-type monoclinic structures.  These latter phases are found
to exhibit huge shear piezoelectric coefficients.

\end{abstract}

\pacs{PACS:77.84.Dy,77.80.Fm,77.65.Bn}

\vskip2pc]

\narrowtext

\marginparwidth 2.7in
\marginparsep 0.5in

The existence of a morphotropic phase boundary (MPB) is an important
feature of the phase diagram of many technologically interesting
perovskite alloys, such as
[Pb(Zn$_{1/3}$Nb$_{2/3}$)O$_{3}$]$_{(1-x)}$[PbTiO$_{3}$]$_x$ (PZN-PT),
[Pb(Sc$_{1/2}$Nb$_{1/2}$)O$_{3}$]$_{(1-x)}$[PbTiO$_{3}$]$_x$ (PSN-PT),
and Pb(Zr$_{1-x}$Ti$_{x}$)O$_{3}$ (PZT) \cite{Park}.  In these
materials, the MPB is the boundary in the temperature-composition
plane separating the rhombohedral phase, in which the electric
polarization is along the [111] direction, from the tetragonal
structure, in which the polarization lies along the [001] direction.
Applying an electric field to these alloys near the MPB generates an
anomalously large change of strain. Very recent theoretical \cite{Fu}
and experimental \cite{Beatriz} studies strongly suggest that such
large piezoelectric response is driven by the electric-field induced
{\it rotation} of the electrical polarization ${\bf P}$, rather than
an electric-field induced change of magnitude of this polarization.

Despite these recent advances, many features related to the effects of
electric fields on structural and piezoelectric properties of
perovskite alloys near the MPB remain puzzling.  For instance, based
on calculations performed on the simple BaTiO$_{3}$ system,
Ref. \cite{Fu} proposed that applying an electric-field along the
pseudo-cubic [001] direction in rhombohedral PZN-PT induces a rotation
of the polarization occurring along the rhombohedral--tetragonal path
(i.e., ${\bf P}$ continuously rotates from the [111] to [001]
direction).  On the other hand, Ref. \cite{Beatriz} suggests a more
complex mechanism in the same PZN-PT system in which the polarization
first follows the rhombohedral--tetragonal path for small electric
field, and then jumps to a new path connecting an orthorhombic to the
tetragonal structure. However, due to experimental restrictions, the
authors of Ref.  \cite{Beatriz} were only able to observe the second
(and new) polarization path.  As a result, the following questions are
currently unanswered: If there is indeed a new polarization path at
higher electric-field, what is the transitional mechanism allowing
such a change of polarization path?  Is this transitional mechanism
sudden or does it happen over a broad range of electric field?
Another aspect that is poorly understood is the striking difference
between the electric-field behavior on opposite sides of the MPB:
applying an electric field to a tetragonal alloy leads to a much
smaller change of strain than is observed for a rhombohedral
composition \cite{Park}.  Furthermore, most studies investigating
piezoelectricity in perovskite materials focused on the $d_{33}$
piezoelectric coefficient.  As a result, other coefficients that may
exhibit even larger values could have been overlooked. On may also
wonder what is the consequence of the polarization paths proposed in
Ref. \cite{Beatriz} on piezoelectricity.

In this Letter we address the problems summarized
above by investigating the effect of
an electric field on the
structural and piezoelectric properties of PZT alloys near the
MPB.  Starting with a composition that is tetragonal at zero field,
we find that applying an electric field along the [111]
pseudo-cubic direction leads to a polarization that simply rotates
in the ($\bar{1}$10) plane from [001] to [111], i.e., from the
tetragonal (T) to the rhombohedral (R) structure, via an intermediate
A-type monoclinic (M$_{\rm A}$) structure \cite{Noheda1}.
The resulting $d_{33}$ piezoelectric coefficient
is only large very close to the M$_{\rm A}$--R transition.
On the other hand, starting with a rhombohedral PZT composition,
we find that applying an electric field along the [001] pseudo-cubic
direction does {\it not} simply generate the reverse transition
sequence R--M$_{\rm A}$--T.
Instead, the polarization begins to follow this
($\bar{1}$10)-plane R--M$_{\rm A}$--T path at low fields, but before
reaching the T phase, the polarization crosses over to join the
(100)-plane O--M$_{\rm C}$--T path connecting the orthorhombic
(O) to the tetragonal structure via an intermediate C-type monoclinic
(M$_{\rm C}$) structure (as hinted for PZN-PT in Ref.~\onlinecite{Beatriz}).
(The notation for monoclinic phases is that of Ref.~\cite{DavidMorrel},
with M$_{\rm A}$ and M$_{\rm C}$ phases having polarization lying
in the ($\bar{1}$10) and (100) planes, respectively.)
The crossing from the M$_{\rm A}$ to the M$_{\rm C}$ structure
occurs via an intermediate triclinic (Tri) structure that is
stable over a substantial electric-field range.
The overall transition sequence is thus
R--M$_{\rm A}$--Tri--M$_{\rm C}$--T.
We find that both the triclinic and M$_{\rm C}$ phases exhibit
huge shear $d_{15}$ piezoelectric coefficients.

We use the numerical scheme proposed in Ref. \onlinecite{Paper1},
which involves the construction of an alloy effective Hamiltonian
($H_{\rm eff}$) \cite{ZhongDavid} from first-principles calculations
 \cite{LDA,LaurentDavid3,USPP} and its use in
Monte-Carlo simulations to compute finite-temperature properties of
PZT alloys.  The effect of an external static electric field ${\bf E}$
on physical properties is included by adding a $-\bf P\cdot E$ term in
$H_{\rm eff}$ (see Ref.~\cite{Alberto2}).  In the present study, the
temperature is kept fixed at 50\,K, and well-converged results are
obtained with $10\times 10\times 10$ simulation boxes (5000
atoms)~\cite{Paper2}.  We use atomic configurations mimicking maximal
compositional disorder, consistent with experimental conditions
\cite{Cross}.  Data from the simulations on the local
mode vectors ${\bf u}$ (directly proportional to the electric
polarization) and the homogeneous strain tensor 
are used within the
correlation-function approach of Refs~\cite{Alberto2,Alberto1} to
obtain the piezoelectric response.
Up to 2$\times$10$^{5}$ Monte-Carlo sweeps are first performed to equilibrate
the system, and then 2$\times$10$^{5}$ sweeps are used to get the
various statistical averages.

Previous studies \cite{Paper1,Paper2,PaperPSN} have demonstrated
that our effective Hamiltonian approach gives a very good account
of experimental and direct first-principles findings in perovskite
alloys.  In particular, it successfully predicts the existence
of three low-temperature ferroelectric phases for
Pb(Zr$_{1-x}$Ti$_{x}$)O$_{3}$ solid solutions near its MPB
\cite{Paper1}:  a tetragonal (T) phase for larger $x$ compositions
($x~>$ 49\% at 50\,K), a rhombohedral (R) phase for smaller
$x$ compositions ($x~<$ 47.5\% at 50\,K), and the recently
discovered monoclinic (M$_{\rm A}$) phase \cite{Noheda1} in between.
The polarization is parallel to the pseudo-cubic [001], [111],
or [$vv$1] ($0<v<1$) direction for the T, R, or M$_{\rm A}$ phase,
respectively.  The electrical polarization can thus be viewed as
rotating in a ($\bar{1}$10) plane from [001] to [111] as the Ti
composition decreases in the monoclinic phase \cite{Paper1}.
(In the following, we  adopt the convention that
the $x$, $y$ and $z$ axes are chosen along the pseudo-cubic
[100], [010] and [001] directions, respectively, and the
piezoelectric tensor is represented in the orthonormal basis
formed by {\bf a$_{1}$} = [100], {\bf a$_{2}$} = [010] and {\bf
a$_{3}$} = [001].)

We first consider a tetragonal PZT composition, $x$=0.5, under an
electric field applied along the [111] direction.  The polarization of this
alloy is initially along $\hat{z}$ when no electric field is applied.
Figure~1(a) displays the cartesian components ($u_{x}$, $u_{y}$
and $u_{z}$) of the supercell average of the local mode vectors
in  this solid solution as a function of the magnitude of the
electric field. It demonstrates that applying a small electric field
along the [111] direction in Pb(Zr$_{0.50}$Ti$_{0.50}$)O$_{3}$
perturbs the system in the expected way: the polarization rotates
away from the [001] direction,
as evidenced by the fact that $u_{x}$ and $u_{y}$ become non-zero.
The strain tensor given by our simulations indicates that the
resulting structure is the monoclinic M$_{\rm A}$ structure.  As the
magnitude of the electric field along the [111] direction increases,
$u_{z}$ decreases, while $u_{x}$ and $u_{y}$ increase and remain
equal.  The polarization in this M$_{\rm A}$ structure
thus rotates towards the [111] direction within the ($\bar{1}$10)
plane.  For a sufficiently large electric field
($\sim$110$\,\times\,10^6\,$V/m), $u_{z}$ becomes equal to $u_{x}$
and $u_{y}$, signaling a phase transition to a rhombohedral
structure \cite{explan-sym} with the polarization aligned along
the [111] direction.
The path followed by the polarization in going from the T via the
M$_{\rm A}$ to the R structure with increasing field is illustrated
in the insert of Fig.~1(a).

\begin{figure}
\epsfxsize=\hsize\epsfbox{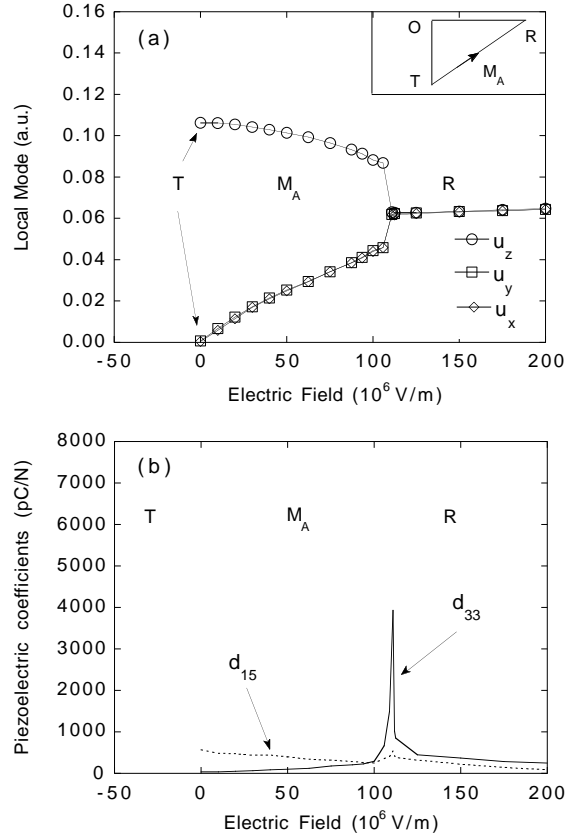}
\caption{(a) Cartesian components $u_{x}$, $u_{y}$ and $u_{z}$
of the local-mode vector, and (b) $d_{33}$ and $d_{15}$
piezoelectric coefficients, as a function of the magnitude
of the electric field applied along the pseudo-cubic [111] direction
in disordered single crystals of Pb(Zr$_{0.50}$Ti$_{0.50}$)O$_{3}$
at T=50\,K. Insert illustrates the path followed by the polarization.}
\end{figure}

Figure 1(b) shows the resulting
behavior of the $d_{33}$ and $d_{15}$ piezoelectric coefficients. 
We decided to focus on these two coefficients because we numerically
found that they are the largest ones among all the 18 piezoelectric
constants. One can note  that $d_{15}$ is larger than $d_{33}$ in the
T structure (and in most of the range of stability of the M$_{\rm A}$ structure).
This is consistent
with recent measurements revealing that the piezoelectric elongation
of the tetragonal unit cell of PZT does not occur along the polar
[001] direction \cite{Noheda3}.  The large value of  $d_{15}$
also explains the strong (averaged) piezoelectric response observed in
ceramic tetragonal samples~\cite{Paper1}.  One can also see that $d_{33}$ is predicted to
be larger in the R phase than in the T and
M$_{\rm A}$ structures. This prediction is consistent with the findings
of Refs.~\cite{Park,Du} that the $d_{33}$ coefficient of rhombohedral
materials can be very large along the [001] direction when the
polar direction is oriented along the pseudo-cubic [111] direction.
Figure~1(b) also reveals that $d_{33}$ adopts values larger than
1,000 pC/N in a narrow electric-field range centered around
the M$_{\rm A}$--to--R transition. Note that $d_{15}$
is also enhanced in this region, but has much smaller values than
$d_{33}$.

We now investigate the effect of an electric field applied along
the [001] direction to a rhombohedral PZT composition,
Pb(Zr$_{0.53}$Ti$_{0.47}$)O$_{3}$.  The polarization
of this alloy starts along the pseudo-cubic [111] direction (i.e,
$u_{x}=u_{y}=u_{z}$) when no external electric field is present.
Figure 2(a) reveals an unusual behavior of the
polarization and a resulting rich phase diagram.  As expected,
applying a small electric field along the [001] direction
of Pb(Zr$_{0.53}$Ti$_{0.47}$)O$_{3}$ perturbs the R structure to
the M$_{\rm A}$ structure, for which $0 < u_{x}=u_{y} < u_{z}$.  
As the
magnitude of the electric field increases, the polarization
rotates towards the [001] direction, with $u_{z}$ increasing while
$u_{x}$ and $u_{y}$ are decreasing.  However, as shown in Fig.~2(a),
Pb(Zr$_{0.53}$Ti$_{0.47}$)O$_{3}$  then adopts two behaviors that
do not occur in the Pb(Zr$_{0.50}$Ti$_{0.50}$)O$_{3}$ alloy.
First, for an electric field ranging between 50 and
88$\,\times\,10^6\,$V/m, $u_{x}$, $u_{y}$  and $u_{z}$ are all non-zero
and different from each other, and the strain tensor identifies
this as a {\it triclinic} (Tri) phase.
Second, for an electric field ranging between 88 and
110$\,\times\,10^6\,$V/m, $u_{x}$ vanishes while $u_{y}$ and
$u_{z}$ remain zero and different from each other.  This
characterizes a second type of monoclinic phase denoted M$_{\rm
C}$ \cite{DavidMorrel,note2,note}.  At larger electric field,
Pb(Zr$_{0.53}$Ti$_{0.47}$)O$_{3}$ finally transforms from this
monoclinic (M$_{\rm C})$ phase into the tetragonal (T) phase,
for which the polarization is along the [001] direction ($u_{z}$
is then the only nonzero component).

The insert of Fig.~2(b) summarizes the path followed by
the polarization in Pb(Zr$_{0.53}$Ti$_{0.47}$)O$_{3}$:
$\bf P$ first rotates in the ($\bar{1}$10) plane from [111] 
to [vv1] (R to M$_{\rm A}$), then switches via [wv1]
(Tri) to the (100) plane, where it finishes by rotating from
[0v1] (M$_{\rm C}$) to [001] (T).  That is, the polarization
starts along the R--M$_{\rm A}$--T path and then switches
to the O--M$_{\rm C}$--T path to complete the transformation
to the T phase.

The electric-field induced R--M$_{\rm A}$--T transformation
path has been observed experimentally in nominally
rhombohedral PZT alloys \cite{Noheda3}, but the unavailability of
single-crystal PZT complicates the experimental determination of the
polarization paths depicted in Fig.~2(a). Furthermore, we find that
the change of path occurs at electric fields that are high
compared to those that can be used experimentally
(our calculations predict that an electric field of
$\sim100\,\times\,10^6\,$V/m is needed to see the transitions
to the Tri and M$_{\rm C}$ phases, while the highest field
applied experimentally in Ref.~\cite{Beatriz} is 2$\,\times\,10^6\,$V/m).
However, we find that the transformations displayed
in Fig.~2(a) all occur at smaller electric fields in PZT alloys
for which the Ti composition $x$ is closer to the MPB:
increasing the Ti composition in PZT from 46\% to 47\% leads to
a considerable decrease of the critical electric field, giving
rise to the Tri phase at $\sim\,$50 instead of
150$\,\times\,10^6\,$V/m, and to the M$_{\rm C}$ phase at $\sim\,$88
instead of 237$\,\times\,10^6\,$V/m. Thus a practical observation
of the change of polarization path depicted in Fig.~2(a) would
probably require the use of a single-crystal PZT alloy with
a composition lying as close as possible to the MPB.

\begin{figure}
\epsfxsize=\hsize\epsfbox{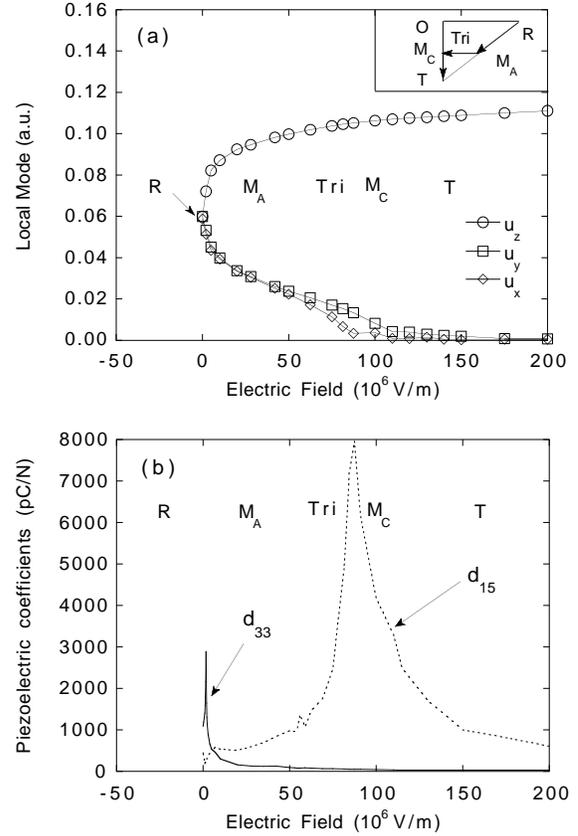}
\caption{(a) Cartesian components $u_{x}$, $u_{y}$ and $u_{z}$
of the local-mode vector, and (b) $d_{33}$ and $d_{15}$
piezoelectric coefficients, as a function of the magnitude
of the electric field applied along the pseudo-cubic [001] direction
in disordered single crystals of Pb(Zr$_{0.53}$Ti$_{0.47}$)O$_{3}$
at T=50\,K. Insert illustrates the path followed by the polarization.}
\end{figure}

While the R--M$_{\rm A}$--T path of the polarization has recently
been predicted to occur in (zero-field rhombohedral) BaTiO$_{3}$
when applying an [001]-oriented electric field, no change of path
has been predicted to occur in this material even at very high
field \cite{Fu,Alberto2}.  Similarly, the O--M$_{\rm C}$--T path
has recently been observed  in (zero-field rhombohedral)
92\%--PbZn$_{1/3}$Nb$_{2/3}$O$_{3}$ 8\%--PbTiO$_{3}$ under an
[001]-oriented electric field \cite{Beatriz}.
However, because of some experimental limitations, the R--M$_{\rm A}$--T
path and the transitional triclinic phase, that probably both
occur at small fields,
have yet to be observed in this material.
To our knowledge, the
present study is thus the first demonstrating the existence
of these two different polarization paths, and the transitional
phase connecting them, in a cubic perovskite system under electric
field \cite{noteAlberto}.

Furthermore, Fig.~2(b) clearly shows that these unusual polarization
paths have a drastic effect on the $d_{33}$ and $d_{15}$
piezoelectric coefficients of Pb(Zr$_{0.53}$Ti$_{0.47}$)O$_{3}$.
For instance, Fig.~2(b) reveals that the $d_{33}$  coefficient
suddenly peaks at a small field, when
Pb(Zr$_{0.53}$Ti$_{0.47}$)O$_{3}$ adopts the M$_{\rm A}$
structure.  This peak corresponds to a sudden increase of the
strain $\eta_{3}$.  $d_{33}$ then decreases with a further increase
of the magnitude of the electric field, implying that the strain
$\eta_{3}$ then adopts a flatter strain-{\it vs.}-field slope.
Interestingly, both the sudden increase and the flatter behavior
of the strain have been observed in rhombohedral PZN-PT \cite{Park}
and were predicted to occur in rhombohedral BaTiO$_{3}$ \cite{Fu}
when applying an [001]-oriented electric field.

The most striking result of Fig.~2(b) is the huge enhancement
of the shear $d_{15}$ piezoelectric coefficient associated with
the change of  polarization  path: $d_{15}$
increases from 1,000 to 8,000\,pC/N with increasing electric
field within the transitional Tri phase.  Once the
polarization is along the O-M$_{\rm C}$-T path (i.e., when the
system has transformed to the M$_{\rm C}$ phase), $d_{15}$
decreases with a further increase of the electric field \cite{note2}.
However, $d_{15}$ remains large, even in the tetragonal phase
under very high electric field. As a result, $d_{15}$ is larger
than 2,000\,pC/N for a wide range of electric fields,
e.g., $\sim\,$50$\,\times\,10^6\,$V/m.

In summary, we have used the first-principles derived computational
scheme proposed in Ref.~\cite{Paper1} to study the structural and
piezoelectric properties of disordered Pb(Zr$_{1-x}$Ti$_{x}$)O$_{3}$
solid solutions near the MPB under an electric field.  Our main
results are: (1) the response of the polarization to an
electric field is qualitatively different in the tetragonal and
rhombohedral phases; (2) specifically, rhombohedral alloys near the
MPB exhibit an unusual switching of the polarization path under
applied field; (3) a large piezoelectric response can result from such
a path switch; and (4) the shear piezoelectric coefficients
(e.g., $d_{15}$) could be exploited to achieve a
drastic improvement in response in piezoelectric devices.

L.B.~thanks the financial assistance provided by the Arkansas
Science and Technology Authority (grant N99-B-21), the Office of
Naval Research (grants N00014-00-1-0542 and N00014-01-1-0600)
and the National Science Foundation (grant DMR-9983678). 
A.G.\ acknowledges support from
the Spanish Ministry of Education (grant PB98-0244).
D.V.\ acknowledges ONR grant N00014-97-1-0048.  We wish to  thank
B. Noheda and W.F. Oliver for very useful discussions.


\begin{references}
\bibitem{explan-sym}
For symmetry reasons, this is expected to be a first-order transition.
\bibitem{note2} Note that a close inspection of the symmetry of the
piezoelectric tensor confirms the identification of the Tri and
M$_{\rm C}$ phases.
\bibitem{note} The case $u_{z} > u_{y} > u_{x}$ is
equivalent by symmetry to the case $u_{z} > u_{x}
> u_{y}$ for both the triclinic (Tri) and monoclinic (M$_{\rm C}$)
phases. Our simulation led to the first case by ``accident,''
i.e., it could also have generated the second situation. Note
that the piezoelectric coefficient $d_{15}$ shown in Fig.~2(b) is
consistent with the first case, i.e. the `1' index corresponds
to an electric field applied along the x-axis -- which has the
smallest cartesian component of the local mode -- while `5'
refers to the shear $\eta_{xz}$ strain -- i.e., coupling the axes
exhibiting the smallest and largest cartesian components of the
local mode.
\bibitem{noteAlberto} It should be noted that we find that
the polarization takes the same complex path, in
reverse, upon the progressive release of the field. This reversibility
might be intrinsic to a defect-free system and has not been observed
experimentally \cite{Beatriz}.
\end{references}
\end{document}